\documentclass[aip,reprint]{revtex4-1}
\usepackage{graphicx}
\usepackage{xcolor}

\draft 

\begin{document}


\title{Optical Coherence Tomography with a nonlinear interferometer in the high parametric gain regime} 

\author{Gerard J. Machado}
 \email{gerard.jimenez@icfo.eu.}
 \affiliation{ 
ICFO - Institut de Ciencies Fotoniques, The Barcelona Institute of Science and Technology, Av. Carl Friedrich Gauss 3, 08860 Castelldefels (Barcelona), Spain
}%
\author{Gaetano Frascella}%
\affiliation{ 
Max-Planck Institute for the Science of Light, Staudtstr. 2, Erlangen D-91058, Germany
}%
\affiliation{ 
University of Erlangen-Nuremberg, Staudtstr. 7/B2, 91058 Erlangen, Germany
}%

\author{Juan P. Torres}
 \affiliation{ 
ICFO - Institut de Ciencies Fotoniques, The Barcelona Institute of Science and Technology, Av. Carl Friedrich Gauss 3, 08860 Castelldefels (Barcelona), Spain
}%
 \affiliation{ 
Department of Signal Theory and Communications, Universitat Politecnica de Catalunya, 08034 Barcelona, Spain
}%

\author{Maria V. Chekhova}
\affiliation{ 
Max-Planck Institute for the Science of Light, Staudtstr. 2, Erlangen D-91058, Germany
}%
\affiliation{ 
University of Erlangen-Nuremberg, Staudtstr. 7/B2, 91058 Erlangen, Germany
}%

\date{\today}

\begin{abstract}
We demonstrate optical coherence tomography based on an SU(1,1) nonlinear interferometer with high-gain parametric down-conversion. For imaging and sensing applications, this scheme promises to outperform previous experiments working at low parametric gain, since higher photon fluxes provide lower integration times for obtaining high-quality images. In this way one can avoid using single-photon detectors or CCD cameras with very high sensitivities, and standard spectrometers can be used instead. Other advantages are: higher sensitivity to small loss and amplification before detection, so that the detected light power considerably exceeds the probing one.
\end{abstract}

\pacs{Parametric down-conversion, optical coherence tomography, nonlinear interferometers}
\maketitle

\section{Introduction}
\label{sec:intro}
Optical Coherence Tomography (OCT)\cite{Huang1991,dresel} is a 3D optical technique that permits high-resolution tomographic imaging. It is applied in many areas of science and technology~\cite{octnews}, from medicine \cite{OCTbook} to art conservation studies \cite{desmond}. To obtain good transverse resolution (perpendicular to the beam propagation axis), OCT focuses light into a small spot that is scanned over the sample. To obtain good resolution in the axial direction (along the beam propagation), OCT uses light with a large bandwidth to do optical sectioning of the sample.

Standard OCT schemes make use of a Michelson interferometer, where light in one arm illuminates the sample and light in the other arm serves as a reference. In the last few years, there has been a growing interest in a new type of OCT scheme \cite{shapiro2009,valles2018optical,Paterova2018OCT,vanselow2019}  that uses so-called {\em nonlinear interferometers} based on optical parametric amplifiers. The latter can be realized with parametric down-conversion (PDC) in nonlinear crystals or four-wave mixing (FWM) in fibers or atomic systems~\cite{Chekhova2016}.

Some of these OCT schemes \cite{shapiro2009,valles2018optical} are based on the idea of {\em induced coherence} \cite{Zou1991,zou1992}, a particular class of nonlinear interferometer originally introduced the very same year as OCT. A parametric down-converter generates pairs of signal and idler photons. The idler beam is reflected from a sample with reflectivity $r_i$ before being injected into a second parametric down-converter. The signal beams coming from the two coherently-pumped parametric sources show a degree of first-order coherence that depends on the reflectivity.  If not only the idler but also the signal from the first down-converter is injected into the second crystal, the scheme turns into an SU(1,1) interferometer~\cite{Yurke:86}. Some OCT schemes use this configuration~\cite{Paterova2018OCT,vanselow2019}. 

Nonlinear interferometers are key elements in numerous applications beyond OCT, namely in imaging \cite{Barreto2014,cardoso2018}, sensing \cite{kutas2020}, spectroscopy \cite{kalasnikov2016,Paterova2018} and microscopy \cite{Kviatkovsky2020,paterova2020}. From a practical point of view, an advantage of these systems is that one can choose a wavelength for the idler beam, which interacts with the sample and is never detected, and another wavelength for the signal beam, which is detected with a high efficiency. This is why the general term `measurements with undetected photons' is used for such systems. 

Optical parametric amplifiers can work in two regimes. In the low parametric gain regime, the number of photons per mode generated is much smaller than one~\cite{silberberg2005}. Many applications have been demonstrated in this regime~\cite{Barreto2014,kalasnikov2016,valles2018optical,Paterova2018OCT}. In the high parametric gain regime~\cite{iskhakov}, the number of photons per mode is higher than one. Applications for imaging have been considered in both regimes~\cite{boyd} and an OCT scheme based on induced coherence and large parametric gain \cite{shapiro2009} has been demonstrated.

Despite the availability of strongly pumped SU(1,1) interferometers~\cite{chekhova2}, recent experiments in OCT using this scheme~\cite{Paterova2018OCT,vanselow2019} work at low parametric gain. In this scenario, the generated photon pair flux is low, which is detrimental for OCT applications.  Here we demonstrate an OCT scheme that makes use of an SU(1,1) interferometer at high parametric gain, generating thus high photon fluxes.

The regime of high parametric gain has several advantages for sensing and imaging. Higher photon fluxes allow using \textit{conventional} charge-coupled device (CCD) cameras or spectrometers, instead of single-photon detectors, and images with high signal-to-noise ratio can be obtained with shorter acquisition times. Meanwhile, unlike in conventional OCT or in the case of low parametric gain, the detected power is considerably higher than the power probing the sample. The nonlinear dependence of the interference visibility on the idler loss makes this regime more sensitive to small reflectivities. Finally, the high-gain regime provides larger frequency bandwidths~\cite{chekhova3}, which would yield higher resolution in OCT schemes. 

\section{Optical coherence tomography scheme}
\begin{figure}[b!]
\includegraphics[width=\linewidth]{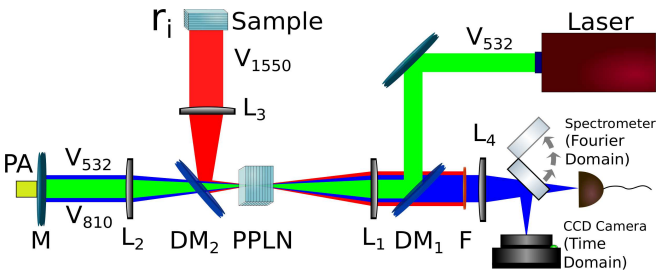}
\caption{Experimental setup for OCT. $DM_{1,2}$ stands for dichroic mirrors, $L_{1,2,3,4}$ are lenses, PPLN is a periodically poled lithium niobate crystal, M is a mirror and F is a short-pass filter. V designates vertical polarization with sub-indexes indicating the wavelengths of the corresponding beams. PA is a piezoelectric actuator. With a flip mirror we can choose to measure the flux rate of signal photons or its spectrum.}
\label{fig:setup}
\end{figure}

In the experiment (Fig.~\ref{fig:setup}), the pump is a Nd:YAG laser generating 18 ps pulses at wavelength $\lambda_{p}=$ 532 nm, with a repetition rate of 1 kHz. The pump illuminates an $L=1$ mm periodically poled lithium niobate (PPLN) crystal and generates signal and idler beams at central wavelengths $810$ nm and $1550$ nm with the same vertical polarization as the pump beam. The bandwidth of the signal spectrum is $8 \pm 1$ nm full-width at half maximum (FWHM) and the bandwidth of the idler is $30 \pm 3$ nm. 

To increase the parametric gain $G$, the pump is focused onto the nonlinear crystal by means of lens $L_{1}$ with the focal length $f_{1}=$ 200 mm. The beam size at the crystal has FWHM $40\pm10$ $\mu$m. The signal and idler beams are separated by long-pass dichroic mirror $DM_{2}$ with the transmission edge at $950$ nm: the pump and the signal are transmitted, while the idler is reflected. The pump and signal beams form the \textit{reference arm} of the interferometer and are collimated using lens $L_{2}$ with the focal length $f_{2}=$ 200 mm. The idler beam constitutes the \textit{probing arm} of the interferometer and is collimated using lens $L_{3}$ with the focal length $f_{3}=$ 150 mm.

The pump and signal beams are reflected by a mirror and both are focused back onto the nonlinear crystal by means of lens $L_{2}$. We consider a reflectivity $r_{s}$ for the signal beam that takes into account losses in optical elements along the signal path. The distance traveled by the signal beam $s_1$ before reaching the nonlinear crystal is $z_{s}$. The idler beam $i_1$ interacts with an object with reflectivity $r_{i}(\Omega)$  and is focused back on the crystal by lens $L_{3}$. It propagates a total distance $z_{i}$. Finally, after parametric amplification in the second pass of the pump by the nonlinear crystal, the signal ($s_2$) and idler ($i_2$) beams are transmitted by the dichroic mirror $DM_{1}$. The idler radiation is filtered out by the short-pass filter F.

A flip mirror allows us to switch the detection between a CCD camera and a spectrometer. In the first case, we reflect signal $s_2$ to the CCD camera (beam profiler) placed in the Fourier plane of lens $L_{4}$ (focal length $f_{4}=$ 100 mm). In this scenario, if the arms of the interferometer are balanced up to the coherence length of PDC radiation, interference fringes appear when the phase of the pump beam is scanned by means of piezoelectric actuator PA.

To measure the spectrum, the signal beam $s_2$ is spatially filtered in the Fourier plane of lens $L_{4}$ and fiber-coupled to a visible spectrometer. In this scenario, spectral interference is observed without scanning the phase, regardless of the optical path difference~\cite{zou1992}. However, it will only be discernible if the interference fringes are resolved by the spectrometer.

OCT requires the use of a large bandwidth ($\Delta_{DC}$) of the PDC spectrum for obtaining high axial resolution of samples. Under the approximation of continuous-wave pump~\cite{dayan2007,chekhova3}, valid also for picosecond pump pulses, the signal beam spectrum $S(\Omega)$  is (see Section I of Supplementary Material)
\begin{eqnarray}
& & S(\Omega)=\big[ 1-|r_i(-\Omega)|^2 \big] \left| V_{s}(\Omega) \right|^2  \nonumber \\
& & + \big| r_s U_{s}(\Omega) V_{s}(\Omega) \exp \left[ i \varphi_{s}(\Omega) \right]  \nonumber \\
& &  + r_i^{*}(-\Omega) U_{i}^*(-\Omega) V_{s}(\Omega) \exp \left[-i \varphi_{i}(-\Omega) \right] \big|^{2},  
\label{spectrum}
\end{eqnarray}
with
\begin{eqnarray}
& & U_{s,i}(\Omega)=  \left\{ \cosh (\Gamma L)-i\frac{\Delta_{s,i}}{2\Gamma} \sinh (\Gamma L) \right\} \nonumber\\
& & \times \exp \left\{ i\left[ k_p+k_{s,i}(\Omega)-k_{i,s} (-\Omega) \right] \frac{L}{2}\right\} ,  \\
& & V_{s,i}(\Omega)= -i \frac{\sigma}{\Gamma}  \sinh (\Gamma L) \nonumber\\
& & \times \exp \left\{ i\left[ k_p+k_{s,i}(\Omega)-k_{i,s} (-\Omega) \right] \frac{L}{2}\right\} \nonumber. 
\label{u,v}
\end{eqnarray}
The subscripts \textit{p,s,i} refer to the pump, signal and idler beams, $k_{p,s,i}$ are wavevectors, $\Omega$ designates the frequency deviation from the central frequencies $\omega_{s,i}$, $\varphi_{s,i}(\Omega)=(\omega_{s,i}+\Omega)z_{s,i}/c$, $\Gamma=(\sigma^2-\Delta_s^2/4)^{1/2}$, the phase-matching functions are $\Delta_s=-\Delta_i=(D_i-D_s)\Omega$ and $D_{s,i}$ are inverse group velocities at the signal and idler central frequencies, respectively. 

The nonlinear coefficient $\sigma$ is
\begin{equation}
\sigma=\left( \frac{\hbar \omega_p \omega_s \omega_i [\chi^{(2)}]^2 R_p}{8\epsilon_0 c^3 A n_p n_s n_i} \right)^{1/2}
\label{sigma}
\end{equation}
where $\omega_{p,s,i}$ are angular frequencies, $n_{p,s,i}$ the corresponding refractive indices, $A$ the effective area of interaction, and $R_p$ the pump photon flux. We estimate $R_p$ as $R_p=E_p/(\hbar \omega_p)/T_0$ where $T_0$ is the pulse duration, and $E_p$ is the energy per pump pulse. 

The parametric gain $G=\sigma L$ is measured  from the dependence of the intensity $I$ of the output radiation (signal or idler) on the input average pump power $P$~\cite{iskhakov}: $I\propto \sinh^{2}(G)$ and $G \propto \sqrt{P}$. In our setup we measure the gain for the first pass by the nonlinear crystal to be $G=1.7 \pm 0.2$. Thus, the total number of idler photons per pulse probing the sample is estimated to be $\sim 13\,000$ ($\sim$ 7 paired photons per mode), with an idler energy per pulse of $1.6$ fJ and a mean power of $1.6$ pW. Meanwhile, the number of signal photons to be detected, after amplification, is as high as $4\times10^5$ per pulse and $4\times10^8$ per second.

\begin{figure}[b!]
\centering\includegraphics[width=\linewidth]{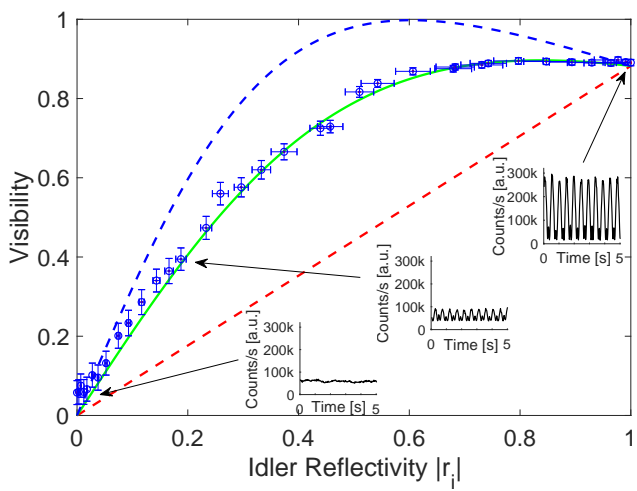}
\caption{Interference visibility as a function of the idler reflectivity. Blue dots are experimental data, error bars are given by the standard deviation, and lines are theoretical predictions. We consider $r_{s}=0.6$. Dashed red curve: $G=0.4$ (low gain regime); green solid line: $G=1.7$; dashed blue line: $G=4.8$. For the theoretical curves we choose the path length difference $\Delta z$ that provides the best visibility.}
\label{fig:vis}
\end{figure}

\section{OCT at high parametric gain: results}
\subsection{Interference visibility at high parametric gain}
To study the dependence of the interference visibility on the reflectivity of the sample, we mimic the latter by a neutral density filter together with a highly reflecting mirror (not shown in Fig. \ref{fig:setup} for simplicity), the total reflectivity coefficient being $r_i$. We minimize the path length difference $\Delta z=z_s-z_i$ between signal and idler beams with mirror M on a translation stage. The phase is scanned with the PA (Fig. \ref{fig:setup}). We measure the flux rate of signal photons $R_s=1/(2\pi)\,\, \int d\Omega\, S(\Omega)$ as a function of $\Delta z$ and determine the visibility $V$ of the fringes according to the definition $V=(R_{max}-R_{min})/(R_{max}+R_{min})$. Here, $R_{max}$  and $R_{min}$ are maximum and minimum of the flux rate, respectively. We repeat this procedure for several values of the reflectivity.

The visibility of interference fringes for multimode radiation in frequency is (Section II of Supplementary Material considers the single-mode approximation)
\begin{equation}
V=\frac{2 |r_s| |r_i| \,|\nu|}{ (1-|r_i|^2)\alpha   + |r_i|^2 \beta + |r_s|^2 \gamma },  \label{visibility1}
\end{equation}
where
\begin{eqnarray}
& & \nu=\int d\Omega\,
U_{s}(\Omega) U_{i}(-\Omega) \big|V_{s}(\Omega) \big|^2 \exp \left\{ i \frac{\Omega}{c} \Delta z\right\} \nonumber, \\
& & \alpha=\int d\Omega\,  \left| V_{s}(\Omega) \right|^2  \nonumber, \\
& & \beta=\int d\Omega\,\left| U_{i}(-\Omega) \right|^2 \,\left| V_{s}(\Omega) \right|^2 \nonumber, \\
& & \gamma=\int d\Omega\,\left| U_{s}(\Omega) \right|^2 \,\left| V_{s}(\Omega) \right|^2. \label{visibility2}
\end{eqnarray}
In order to observe fringes, $|\Delta z|$ should be smaller than the coherence length of PDC $l_c \sim \lambda_i^2/\Delta \lambda_i$, where $\Delta \lambda_i$ is the bandwidth of the idler beam. 

Figure \ref{fig:vis} shows the visibility measured (blue points) for $G=1.7$, and calculated (solid and dashed lines) for gains $G=0.4$ (red), $1.7$ (green) and $4.8$ (blue), with $r_{s}=0.6$. The gain in the experiment can be varied by changing the mean power of the pump laser. The insets show some examples of the interference pattern measured by scanning the phase in a small region ($\sim$ 0.5 $\mu$m) around the point of maximum visibility, well within the coherence length of PDC ($\sim$ 80 $\mu$m). The red dashed line corresponds to the low parametric gain regime. In this regime, $|V_{s,i}(\Omega)|<<1$ and $|U_{s,i}(\Omega)|\sim 1$, which gives a linear dependence of the visibility on the idler reflectivity~\cite{Zou1991}:
\begin{equation}
V= \frac{2|r_s|}{1+|r_s|^2}|r_i|.
\label{vis:low}
\end{equation}

The green line is the case studied in our experiment, and it yields a nonlinear dependence.  This is similar to the case of induced coherence schemes where interference visibility also depends in a nonlinear fashion on the sample reflectivities~\cite{belinsky1992,wiseman2000}. The experimental results are in good agreement with the theory. The maximum visibility measured is $V=90\%$.   

The blue dashed line corresponds to $G=4.8$, the very high-gain regime. In this case the visibility is 
\begin{equation}
V= \frac{2|r_s|}{|r_i|^2+|r_{s}|^2}|r_{i}|.
\label{vis:high}
\end{equation}
If signal and idler losses are equal, $r_i=r_s$, the visibility is equal to 1. Notice that nonlinear relationships have also been observed for configurations where the first parametric down-converter is seeded with an intense signal beam~\cite{milonni2015,cardoso2018}. 

The value $r_s=0.6$ is the estimated reflectivity in the signal path in our setup. Losses in the signal path $s_{1}$ are not only due to the optics (double pass through an uncoated lens and the dichroic mirror), but also most likely due to spatial mode mismatch. 

The slope of the visibility vs reflectivity curve is an indication of the sensitivity of OCT. At low idler reflectivity, which is the case of interest of multiple applications of OCT, one can observe in Fig. 2 that the slope increases with the gain. For the best case of $r_s=1$, the slope goes from 1 for the low parametric gain regime [see Eq. (6)] to 2 for the regime of very high parametric gain [see Eq. (7)]. This last value is also characteristic of standard OCT.

\subsection{Fourier-domain OCT at high parametric gain} 
In this work, we show that we can do Fourier- or spectral-domain\cite{deboer} OCT (FD-OCT) based on a SU(1,1) interferometer. FD-OCT allows faster data acquisition, is more robust since it has no movable elements, and shows better sensitivity than time-domain OCT \cite{choma}.  

FD-OCT analyzes the modulation of the spectrum of the output signal beam and requires a non-zero value of the path length difference $\Delta z$, contrary to the case of time-domain OCT that requires a value of $\Delta z$ close to zero (Section III of Supplementary Material shows a way to evaluate $\Delta z$). When considering a single reflection, the average fringe separation in angular frequency, $2\pi c/|\Delta z|$, needs to be smaller than the bandwidth of parametric down-conversion $\Delta_{dc}$ and larger than the resolution of the spectrometer, $\delta \omega$, so that the modulation is accurately resolved. Making use of $\delta \omega=(2\pi c/\lambda_s^2) \delta \lambda$ and $\Delta_{dc}=(2\pi c/\lambda_s^2) \Delta \lambda_{s}$, $\Delta z$ is constrained by
\begin{equation}
 \frac{\lambda_s^2}{\Delta \lambda_{s}} \ll |\Delta z| \ll \frac{\lambda_s^2}{\delta \lambda}.
 \label{path_diff}
\end{equation}
In our setup we have $\Delta \lambda_{s} = 8$ nm and $\delta \lambda=1.2$ nm, so $82 \,\mu m \ll |\Delta z| \ll 546 \,\mu m$.

\begin{figure}[t!]
\centering\includegraphics[width=0.96\linewidth]{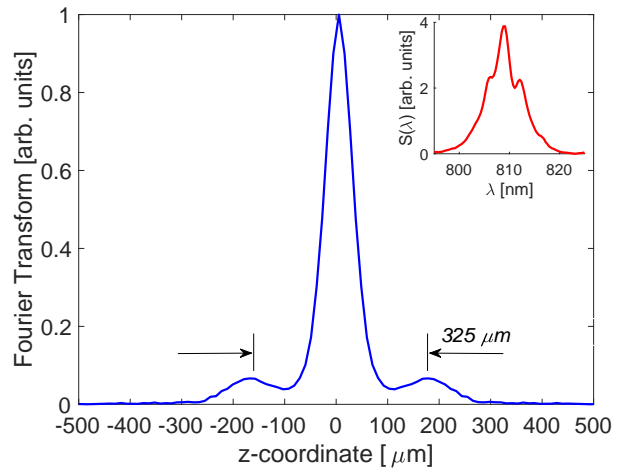}
\caption{Fourier transform of the spectrum measured after re-sampling the spectrum to wavenumbers. The inset shows the measured spectrum $S(\lambda)$. The axial resolution of the OCT scheme (FWHM width of the peaks) is 60 $\mu$m.}
\label{fig:oct}
\end{figure}

The flip mirror at the output of the interferometer is removed to fiber-couple signal photons into a spectrometer. Mirror $M$ is mounted on a nanometric step translation stage to actively control the reference arm length, so that $\Delta z$ is adjusted to an optimum value.  

We probe a $d=100$ $\mu$m thick microscope glass slide with group index $n_g \sim 1.5$, equivalent to a two-layer object with the optical path length $2 n_g d \sim 300$ $\mu$m. The spectrum $S(\lambda)$ is expressed in terms of k-values and Fourier transformed (see Section IV in Supplementary Material for details). $\Delta z_1$ and $\Delta z_2$ are path length differences corresponding to the locations of the two layers and should fulfill Eq. (8). The resolution of the spectrometer ($\delta \lambda$) in our setup does not allow to resolve the sample using a configuration with $|\Delta z_1|, |\Delta z_2| >> 82 \mu$m, which would be the most convenient scenario.

Instead of this we consider several cases where we change the location of the zero path length difference. Figure \ref{fig:oct} shows the case when only three peaks are present in the FT of the spectrum of the signal beam and the distance between the peaks is twice the optical length of the sample (see Section V in Supplementary Material for details). The axial resolution ($\sim 60 \mu$m) can be improved by engineering the phase-matching conditions of nonlinear crystals \cite{nasr,hendrych,abolgamesh,vanselow} and by spectral shaping \cite{tripathi}.

\section{Conclusions \& Outlook}
We have demonstrated a robust and compact OCT scheme based on an SU(1,1) nonlinear interferometer that works in the high gain regime of parametric down-conversion. The setup is versatile since it allows to do both time-domain and Fourier-domain OCT, and can also be easily converted into an induced coherence scheme (see Section VI of Supplementary Material for details). 

The high parametric gain provides high photon fluxes and allows for using standard beam profilers and spectrometers instead of single-photon detectors or highly sensitive CCD cameras. We can easily reach powers of interest for many applications. For instance, in ophthalmology light entering the cornea should have a maximum power of 750 $\mu$W. In art restoration studies typical power is a few milliwatts. The detected power can be even higher, due to the parametric amplification after probing the object. The nonlinear dependence of the interference visibility on the reflectivity makes the high-gain OCT especially sensitive to weakly reflecting samples.

The method still benefits from its well-known salutary features: the sample is probed by near-infrared (NIR) photons, while photodetection takes place in the visible range. This may yield  deeper penetration into samples while measuring at the optimum wavelength with silicon-based photodetectors. In our case, the infrared beam is centered at 1550 nm and the visible at 810 nm. One can generate even bright THz radiation \cite{kitaeva,kutas2020} by clever engineering of nonlinear crystals. Enhanced axial resolution can be achieved by using appropriate parametric down-converters with a broader bandwidth.

\section*{Supplementary Material}
See the Supplementary Material for a complete derivation of Eq. (\ref{spectrum}), the visibility of interference in the single mode case, the measurement of the path length difference $\Delta z$, the Fourier transform analysis of the signal spectrum, and how to transform our setup into an induced coherence scheme. 

\section*{acknowledgments}
We acknowledge support from the Spanish Ministry of Economy and Competitiveness (“Severo Ochoa” program for Centres of Excellence in R\&D, SEV-2015-0522), from Fundacio Privada Cellex, from Fundacio Mir-Puig, and from Generalitat de Catalunya through the CERCA program. GJM was supported by the Secretaria d'Universitats i Recerca del Departament d'Empresa i Coneixement de la Generalitat de Catalunya and European Social Fund (FEDER).

\section*{Data Availability Statement}
The data that support the findings of this study are available from the corresponding author upon request.

\bibliography{references.bib}

\clearpage
\begin{center}
   \bf \huge Supplementary Material  
\end{center}
\setcounter{section}{0}
\setcounter{figure}{0}
\setcounter{equation}{0}
\def\theequation{S\arabic{equation}}
\def\thefigure{S\arabic{figure}}

\section{Derivation of the spectrum of signal photons given by Eq. (1) in the main text.}
The input signal and idler beams are in the vacuum state. The relationship between the input annihilation operators $b_{s}$ and $b_{i}$ and the output operators $a_{s_1}$ and $a_{i_1}$ of signal and idler beams generated after the first pass of the pump pulse by the nonlinear crystal is described by the Bogoliuvov transformation~\cite{navez,brambilla,Torres2011} 
\begin{eqnarray}
& & a_{s_1} (\Omega)=U_{s} (\Omega) b_{s}(\Omega) + V_s(\Omega) b_i^{\dagger}(-\Omega), \nonumber \\
& & a_{i_1}(\Omega)=U_i(\Omega) b_i(\Omega) + V_i(\Omega) b_s^{\dagger}(-\Omega),
\end{eqnarray}
The expressions for $U_s$, $U_i$, $V_s$ and $V_i$ are given in Eq. (2) of the main text. $\Omega$ designates the frequency deviation from the central frequencies of the signal and idler beams, $\omega_{s}$ and $\omega_{i}$. 

The transformations for operators $a_{s_1}$ and $a_{i_1}$ accounting for propagation and loss read
\begin{eqnarray}
& & a_{s_1} (\Omega) \Longrightarrow r_{s}(\Omega) a_{s_1}(\Omega) \exp \left[ i\varphi_s(\Omega) \right]+f_s(\Omega), \nonumber \\ 
& & a_{i_1}(\Omega) \Longrightarrow r_{i}(\Omega) a_{i_1}(\Omega) \exp \left[ i\varphi_i(\Omega) \right]+ f_i(\Omega),
\end{eqnarray}
where $\varphi_{s,i}(\Omega)=(\omega_{s,i}+\Omega) z_{s,i}/c$ and $f_{s,i}$ are operators that fulfill the commutation relations\cite{haus,boyd2} $[f_i(\Omega),f_i^{\dagger}(\Omega^{\prime})]=\left[1-|r_i(\Omega)|^2 \right] \delta(\Omega-\Omega^{\prime})$ and $[f_s(\Omega),f_s^{\dagger}(\Omega^{\prime})]=\left[ 1-|r_s(\Omega)|^2 \right] \delta(\Omega-\Omega^{\prime})$. 
After parametric amplification in the second pass of the pump pulse by the nonlinear crystal, the signal beam $a_{s2}(\Omega)$ reads
\begin{equation}
a_{s_2} (\Omega)=U_{s} (\Omega) a_{s_1}(\Omega) + V_s(\Omega) a_{i_1}^{\dagger}(-\Omega),
\end{equation}
which yields
\begin{widetext}
\begin{eqnarray}
\label{annih_s2}
& & a_{s_2}(\Omega)=\left\{ r_s U_{s}(\Omega) U_{s}(\Omega) \exp \left[ i \varphi_{s}(\Omega) \right] + r_i^{*}(-\Omega) V_{i}^{*}(-\Omega) V_{s}(\Omega)  \exp \left[ -i \varphi_{i}(-\Omega) \right] \right\} b_{s}(\Omega) +  \nonumber \\ 
& &  + \left\{ r_s V_{s}(\Omega) U_{s}(\Omega) \exp \left[ i \varphi_{s}(\Omega) \right] +r_i^{*}(-\Omega) U_{i}^{*}(-\Omega) V_{s}(\Omega) \exp \left[ -i \varphi_{i}(-\Omega) \right] \right\} b_{i}^{\dagger}(-\Omega) +  \nonumber  \\ 
& & + U_{s}(\Omega)f_s(\Omega) + V_{s}(\Omega)f_s^{\dagger}(-\Omega).
\end{eqnarray}
\end{widetext}
We assume that $r_s(\Omega)$ is frequency independent. If we substitute Eq. (\ref{annih_s2}) into $S(\Omega)=\langle a_{s_2}^{\dagger}(\Omega) a_{s_2}(\Omega) \rangle$ we obtain the expression of the spectrum given in Eq. (1) of the main text.

\section{Visibility of interference in the single mode approximation}
Some of the results obtained in the experiments described in the main text can also be well described qualitatively using the single-mode approximation. In this case, the expression for the visibility equivalent to Eq. (4) in the main text is
\begin{equation}
V=\frac{2 |r_s| |r_i| |U_1||U_2||V_1||V_2|}{(1-|r_i|^2) |V_2|^2+|r_i|^2 |U_1|^2 |V_2|^2+|r_s|^2 |U_2|^2 |V_1|^2 },
\end{equation}
where $U_1$ and $V_1$ refer to the first pass by the nonlinear crystal and $U_2$ and $V_2$ to the second pass. If we write $|V_j|=\sinh(G_j)$, $|U_j|=\cosh(G_j)$ and we make use of $|U_j|^2-|V_j|^2=1$ ($j=1,2$), we have
\begin{widetext}
\begin{equation}
V= \frac{|r_s| |r_i|}{2} \frac{\sinh^2(G_1+G_2)-\sinh^2(G_1-G_2)}{\sinh^2(G_2) +|r_s|^2 \sinh^2(G_1)+\left( |r_s|^2+|r_i|^2 \right) \sinh^2(G_1) \sinh^2(G_2)}.
\end{equation}
\end{widetext}
For $r_s=r_i$ we recover Eq. (S1) in Supplementary Material\cite{chekhova2}.

We consider two important limits. In the low parametric gain regime, the values of $G_{1,2}$ are very small, so $\sinh(G_j) \sim G_j$ and $G_1^2 G_2^2 \ll G_1^2,G_2^2$. We thus have 
\begin{equation}
V=\frac{2 |r_s| G_1 G_2}{G_2^2+|r_s|^2 G_1^2}\,\,|r_i|
\end{equation}
that it is the expected linear relationship on $|r_i|$. For $G_1=G_2$ this is Eq. (6) in the main text.

In the very high parametric gain regime, when the difference of gains is small, $\sinh^2 (G_1+G_2) \sim \exp[2(G_1+G_2)]/4 \gg \sinh^2 (G_1-G_2)$ and $\sinh^2 (G_1) \sinh^2(G_2) \sim \exp[2(G_1+G_2)]/16 \gg \sinh^2(G_1), \sinh^2(G_2)$. The visibility is
\begin{equation}
V=\frac{2 |r_s| |r_i|}{|r_s|^2+|r_i|^2}.
\end{equation}
This is Eq. (7) in the main text. Notice two important points for the limit of  very high parametric gain regime, when the difference of gains is not very large: i) The visibility does not depend on the gain, and ii) For $|r_s|=|r_i|$ the visibility is $1$.

\section{Measurement of the path length difference $\Delta z$.}
\begin{figure}[h!]
\centering\includegraphics[width=\linewidth]{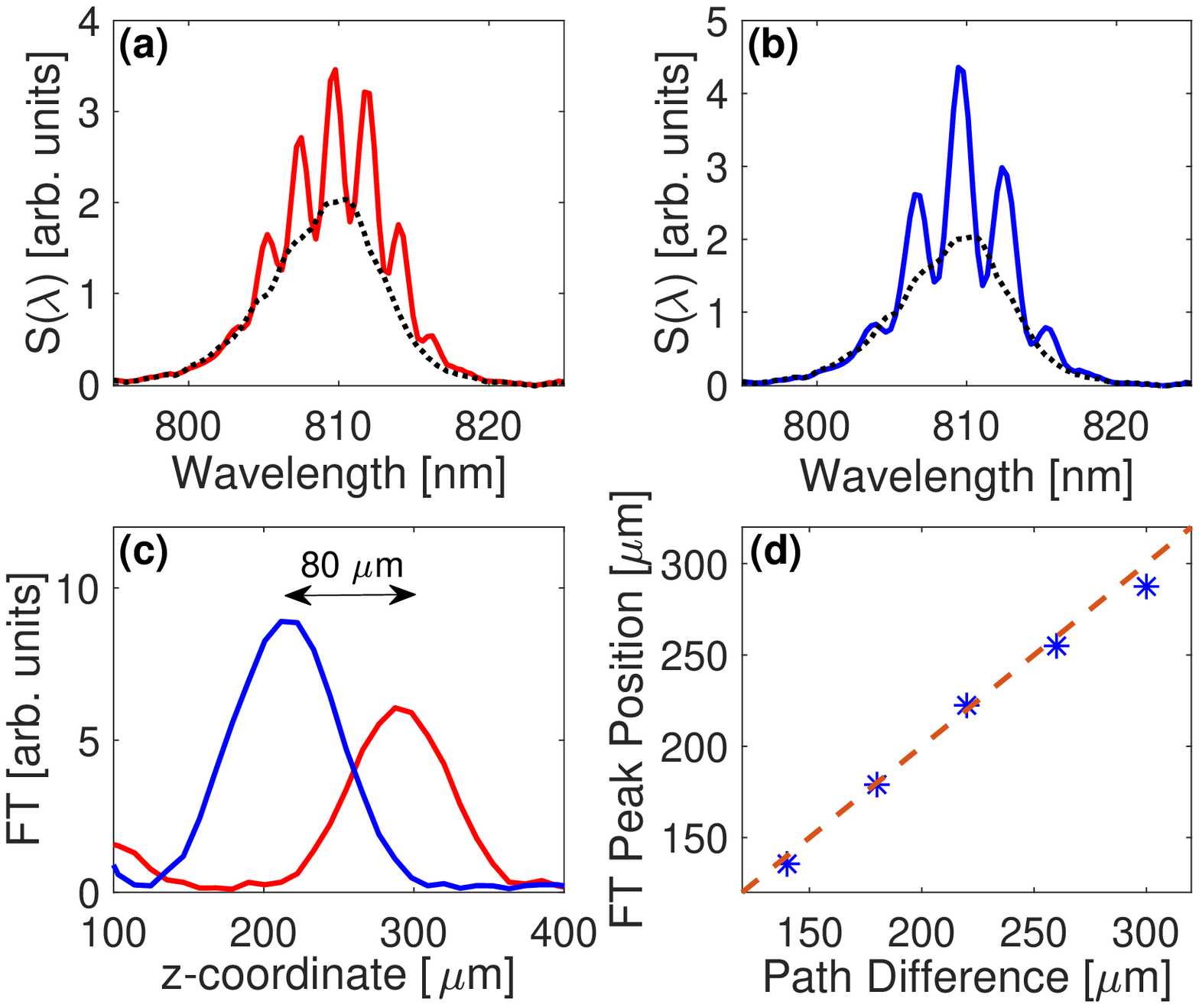}
\caption{\textbf{(a,b)} Spectra measured for two different optical path differences $\Delta z_{1,2}$ (solid line). The dotted lines stand for the spectrum when the idler arm is blocked. \textbf{(a)} $\Delta z_{1} =300$ $\mu$m; \textbf{(b)} $\Delta z_{2} = 220$ $\mu$m. \textbf{(c)} Zoom of the Fourier transforms of the \textbf{(a,b)} spectra after resampling to wavenumbers. The positions of the peaks reveal directly the unbalancing between the arms of the interferometer. The peak separation is 80 $\mu m$, corresponding to the path difference between $\Delta z_{1} - \Delta z_{2}$. \textbf{(d)} Position of the Fourier transform peak versus the path difference. Stars are the experimental data, and the dashed line is the theoretical dependence assuming exact equality.}
\label{fig:deltaz}
\end{figure}

We determine the value of the path length difference $\Delta z$ moving the position of a mirror located in the signal path, and looking at the modulation of the signal beam spectrum as a function of the path length difference.  
Figures \ref{fig:deltaz}(a) and (b) show the spectrum obtained experimentally for two values of the path length difference: $\Delta z_1=220$ $\mu$m and $\Delta z_2=300$ $\mu$m. The visibility of the spectral modulation is affected by the losses in the signal path and by the  resolution of the spectrometer ($\delta \lambda \sim 1.2$ nm). In spite of this, we still can get the information of interest. Figure \ref{fig:deltaz}(c) shows the Fourier transform of the signals shown in Figs. \ref{fig:deltaz}(a) and (b). The separation of the two peaks is 80 $\mu$m, in perfect agreement with the difference between the two values of $\Delta z$ measured. Figure \ref{fig:deltaz}(d) shows the position of the peaks of the Fourier transform as a function of the path length difference $\Delta z$. There is good agreement between theoretical and experimental values.

\section{Fourier Transform analysis}
\begin{figure}[h!]
\centering\includegraphics[width=\linewidth]{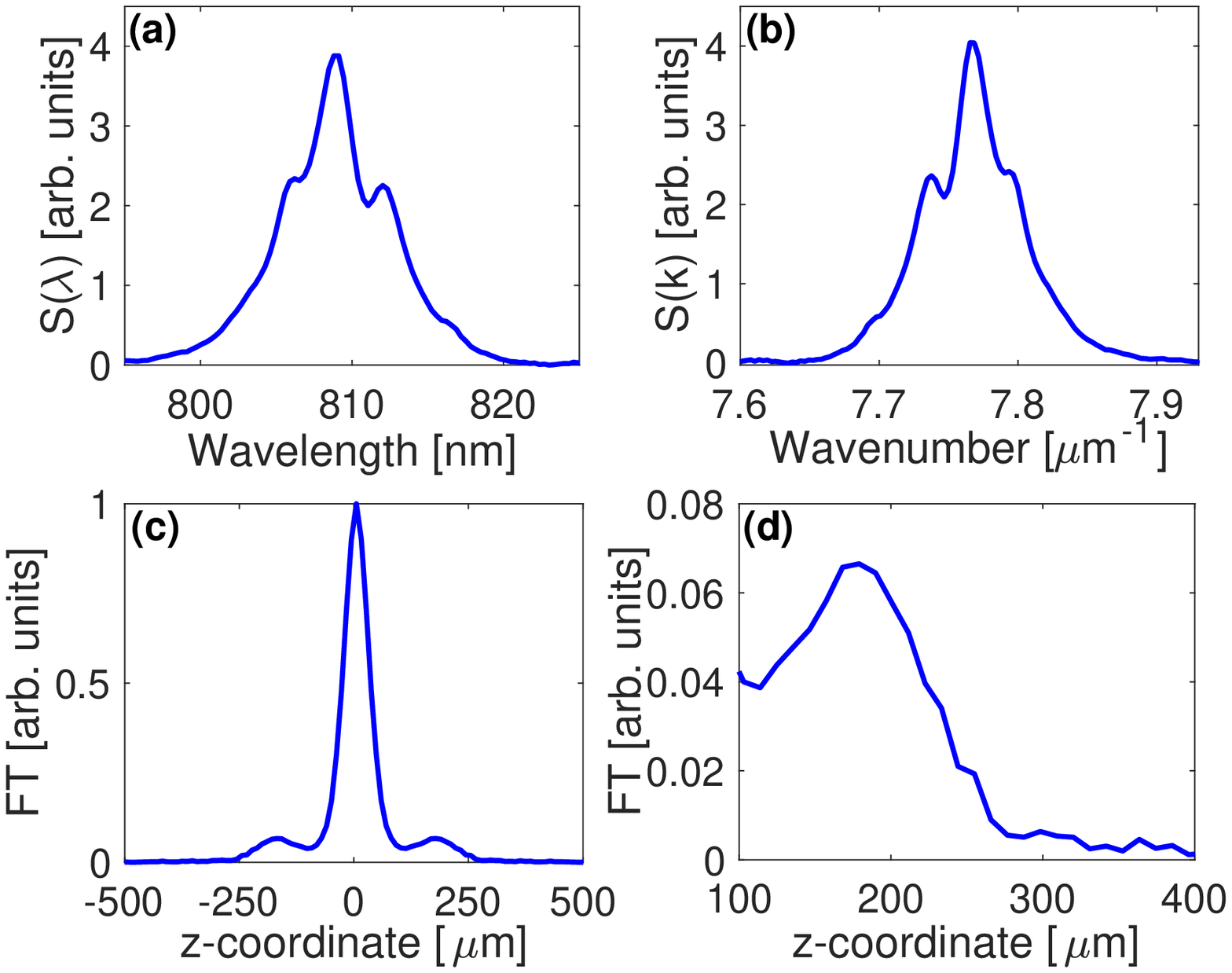}
\caption{Step-by-step procedure to obtain the depth profile of the OCT sample. \textbf{(a)} Spectrum measured with a visible spectrometer, $S(\lambda)$.  \textbf{(b)} Spectrum re-sampled to wavenumbers, $S(k)$.  \textbf{(c)} Fourier transform of the re-sampled spectrum. \textbf{(d)} Zoom of the Fourier transform showing the peak at a positive value of the $z$-coordinate.}
\label{fig:fourier}
\end{figure}

Here we explain in detail how to obtain the Fourier transform of the spectrum measured as a function of the wavelength of the signal beam. Figure \ref{fig:fourier} depicts the step-by-step procedure. 

Figure \ref{fig:fourier}(a) shows the spectrum $S(\lambda)$ measured with a spectrometer sensitive in the visible range. The spectrum is rewritten as function of the wavenumber $k=2\pi/\lambda$. Considering the Jacobian of the transformation, the spectrum is 
\begin{equation}
S(k)=\frac{2\pi}{k^2} S(\lambda)
\end{equation}
The spectrum is re-sampled to obtain a function $S(k)$ with equally-spaced k-values [Fig. \ref{fig:fourier}(b)]. The Fourier transform ($\sim \int dk S(k) \exp(ikz)$) of the re-sampled spectrum is shown as function of the axial position $z$ in Fig. \ref{fig:fourier}(c). Finally, we show a zoom of the FT showing only the peak of the FT at the positive value of the z-coordinate [Fig. \ref{fig:fourier}(d)]. 

\section{The shape of the spectrum of the signal}
In this section we explain why the Fourier transform of the spectrum of the signal beam shown in Figure 3 of the main text shows three peaks, and why the distance between peaks located at $z \ne 0$ is $2n_g d$, twice the optical length of the sample.

Let us define $t$ as the distance from the first layer of the sample to the $z$-position with zero path length difference ($z_s=z_i$). If $t=0$ the position of the first layer is such that $\Delta z_1=0$. If $t=n_g d$, the position of the second layer fulfills $\Delta z_2=0$. If $t<0$ the position of zero path length difference is located before the first layer of the sample, and if $t> n_g d$ the position of zero path length difference is located beyond the second layer of the sample.

\begin{figure}[t!]
\centering\includegraphics[width=\linewidth]{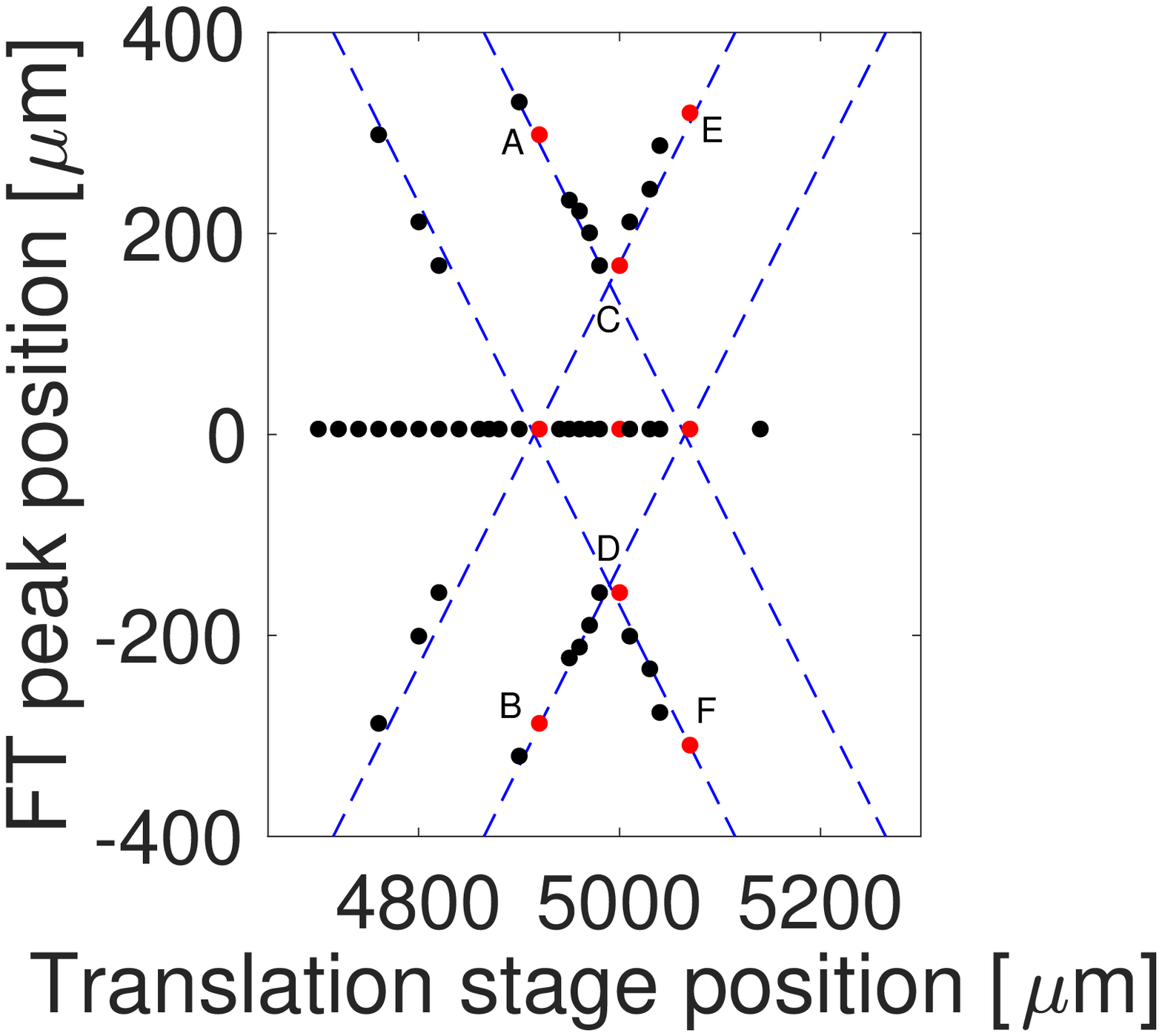}
\caption{Position of the peaks of the Fourier transform of the spectrum of the signal beam as a function of the position of the translation stage. Black dots represent experimental points. The dashed blue lines are theoretical lines based on Eq. (S10) that best fit the experimental data. In this way, we can determine to which value of \textit{t} corresponds each position of the translation stage. Points of interest are indicated as \textit{A, B, C, D, E and F} and highlighted in red.}
\label{newfigure}
\end{figure}

We assume that the shape of the spectrum in the low and high parametric gain regimes is qualitatively similar, while in the low parametric gain regime one can obtain useful analytical expressions of the spectrum $S(k)$ as a function of the wavenumber $k=\Omega/c$. The spectrum is
\begin{eqnarray}
& & S(k)=2\left| V_{s}(k)\right|^2  \left\{ 1+r_1  \cos \left[ \varphi_1+ k \left( c D L +2 t \right) \right] \right. \nonumber \\
& & \left. + r_2 \cos \left[ \varphi_2+k \left( c D L +2 t-2 n_g d \right) \right] \right\}.
\end{eqnarray}
The reflectivities of the first and second layers of the sample are $r_{1,2}$, $D=D_{s}-D_{i}$ where $D_{s,i}$ are inverse group velocities at the signal and idler central frequencies, $n_{g}$ is the group index of the sample and $\varphi_{1,2}$ are constant phases given by
\begin{eqnarray}
& & \varphi_1=\frac{\omega_s}{c} n_s L+\frac{\omega_i}{c} n_i L + 2 \frac{\omega_s}{c} z_s+ 2 \frac{\omega_i}{c} z_{i}  \nonumber \\
& & \varphi_2=\varphi_1+ 2\frac{\omega_i}{c}\, n_0 d.
\end{eqnarray}
$L$ is the length of the nonlinear crystal, $n_0$ is the refractive index of the sample, $z_{s}$ is the distance from the nonlinear crystal to the reference mirror located in the signal arm and $z_{i}$ is the distance from the crystal to the first layer of the sample.

Let us first neglect the effect of the axial resolution for unveiling all peaks of the FT. The first term in Eq. (S10) will produce a peak of the FT at $z=0$. The second term would generate two symmetric peaks at $2t+cDL$ and $-2t-cDL$. Finally the third term would generate peaks at $2t+cDL-2n_g d$ and $-2t-cDL+2n_g d$. We show in Figure (S3) the location of these peaks (dashed lines) and the positions of peaks obtained experimentally (dots). For $t < -cDL/2$ or $t > n_g d-cDL/2$, there will be always five FT peaks and the distance between the two peaks peaks at $z > 0$, or $z<0$, will be $2 n_g d$. If one now considers the effect of the axial resolution for distinguishing the FT peaks, the ideal scenario for OCT is $t \ll -cDL/2$ or $t \gg n_g d-cDL/2$. Unfortunately the sensitivity of the spectrometer used in our experiments impedes us to work in this regime and the axial resolution ($\sim$ 60 $\mu$m) makes the observation of the FT peaks close to the central peak troublesome.

However, for $-cDL/2 < t < n_g d-cDL/2$ we still can determine the distance between layers.  In this scenario the number of FT peaks and the distance between them might change. We focus on three cases that correspond to three values of \textit{t} where only two FT peaks at $z \ne 0$ are expected. In Fig. S3 these points are the ones where the theoretical lines intersect. The first case is for $t=-cDL/2$ (points A and B) and the separation between these peaks is $4 n_g d$. The second case is for $t=n_g d-cDL/2$ (points E and F) and the separation between the peaks is the same. Fig. 3 of the main text corresponds to the third case, that is $t=(n_g L-cDL)/2$ (points C and D). The separation between these peaks is $2 n_g d$.

\vspace{1.5em}
\section{How to transform the SU(1,1) interferometer into an interferometer based on induced coherence}
The experimental setup depicted in Fig. 1 of the main text corresponds to an OCT system based on an SU(1,1) interferometer. We show here how we can easily transform this experimental scheme into an OCT system that makes use of the concept of induced coherence.

\begin{figure}[h!]
\begin{minipage}{.75\linewidth}
\centering
{\includegraphics[width=\linewidth]{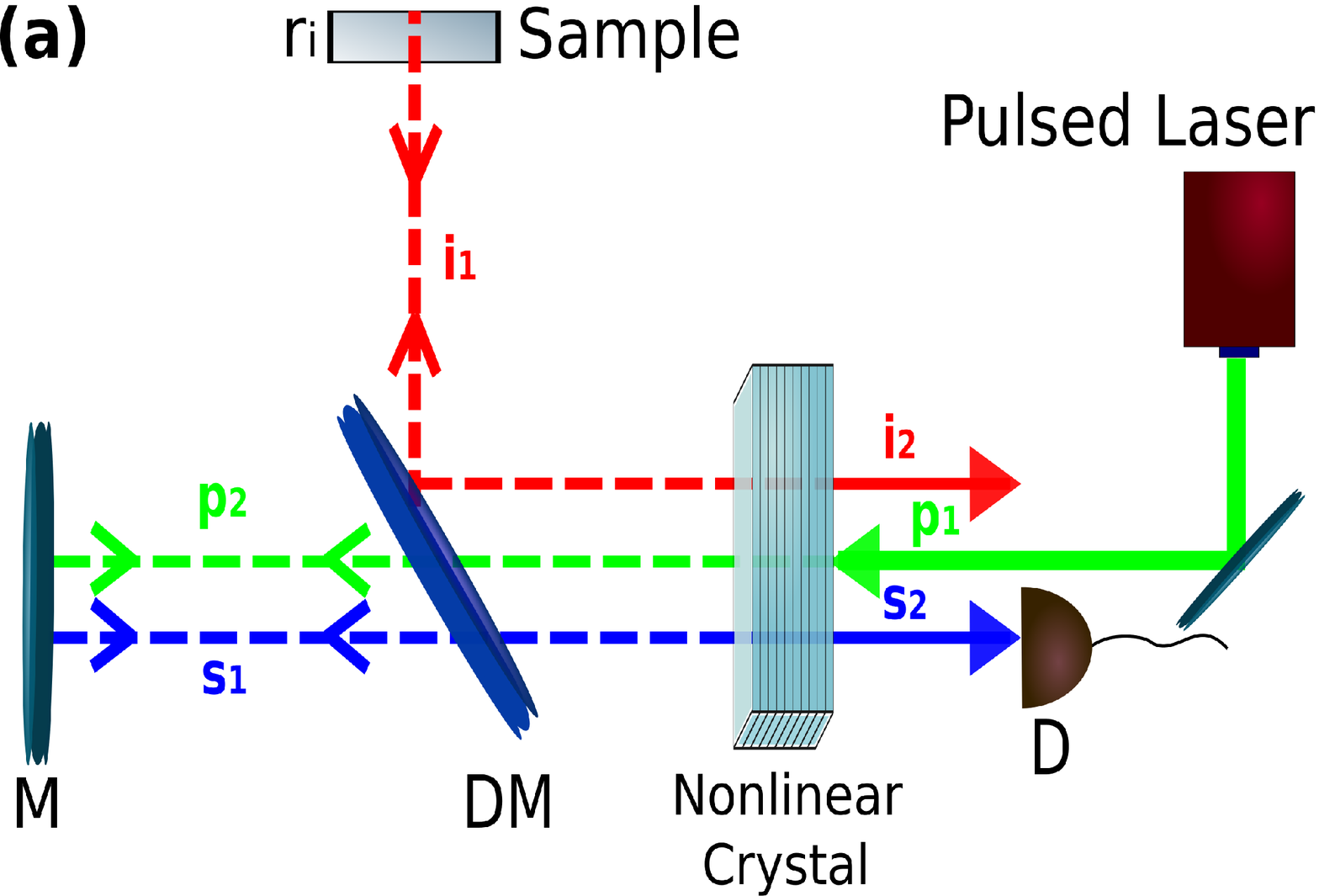}}
\end{minipage}
\begin{minipage}{.75\linewidth}
\centering
{\includegraphics[width=\linewidth]{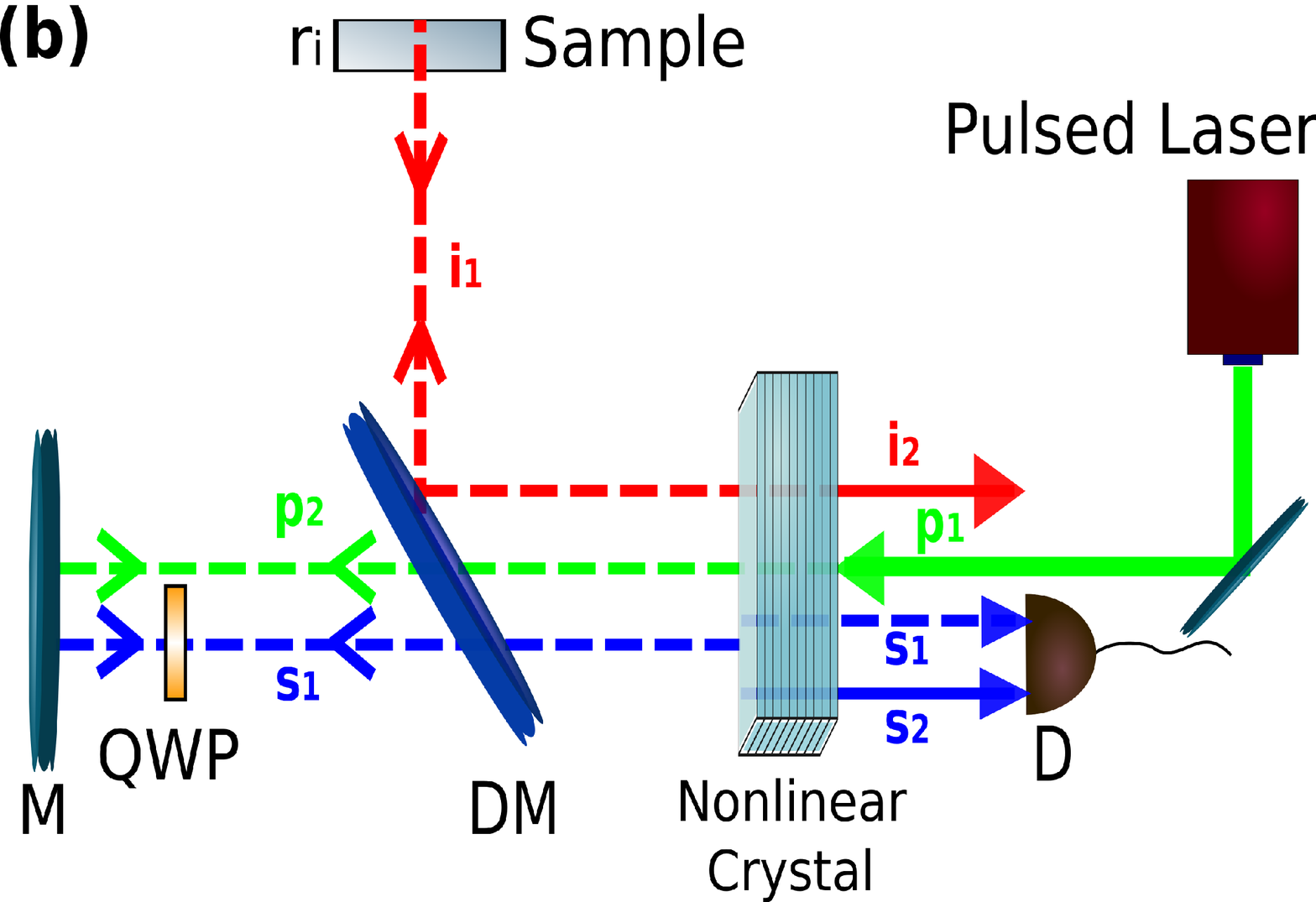}}
\end{minipage}
\caption{\textbf{(a)} Sketch of the experimental setup corresponding to an SU(1,1) interferometer described in the main text. \textbf{(b)} Sketch of a new experimental setup based on the idea of induced coherence.  The pump beams are represented by green lines, the signals are represented by blue lines and the idlers by red lines. A quarter-wave plate QWP placed into the signal beam prevents it from being amplified on the second pass through the nonlinear crystal.}
\label{fig:su11vsind}
\end{figure}

To change the experimental setup from an SU(1,1) interferometer [Fig. \ref{fig:su11vsind}(a)] to a scheme based on the induced coherence effect, one should prevent the signal wave from being amplified on the second pass through the nonlinear crystal. This can be done [see Fig. \ref{fig:su11vsind}(b)] by changing the polarization of signal beam $s_1$ to an orthogonal one with the help of a quarter-wave plate (QWP). Then, only the idler beam $i_1$ would seed the parametric amplification process in the second pass through the nonlinear crystal. In this case, one would distinguish three beams at the output of the nonlinear crystal: idler $i_2$ and two signal beams with orthogonal polarizations: $s_1$ and $s_2$. The detection stage would measure coherence induced between signal beams $s_1$ and $s_2$. Importantly, the QWP should not affect the pump polarization  (be a full-wave plate for the pump).

Finally, we point out that if we change to an orthogonal polarization both the signal and idler beams before the second pass by the nonlinear crystal, without changing the polarization of the pump beam, the set-up can be used to generate polarization entangled photons with high efficiency in the low parametric gain regime \cite{fabian}.

\end{document}